\documentclass{iopart}

\usepackage{amsmath}
\usepackage{iopams}
\usepackage[dvips]{graphicx}
\usepackage[latin1]{inputenc}

\def\lta{~\raise.4ex\hbox{$<$}\llap{\lower.6ex\hbox{$\sim$}}~}
\def\gta{~\raise.4ex\hbox{$>$}\llap{\lower.6ex\hbox{$\sim$}}~}
\def\ie{{\it i.e.}~}

\begin{document}

\title{The Clumping Transition in Niche Competition: \\
a Robust Critical Phenomenon}

\author{H. Fort$^1$\footnote{\mailto{hugo@fisica.edu.uy}}, M. Scheffer$^2$
and E. van Nes$^2$}

\address{$^{1}$Complex Systems Group, Instituto de Física, Facultad de Ciencias,
Universidad de la Rep\'ublica, Igu\'a 4225, 11400 Montevideo,
Uruguay.\\
$^{2}$Wageningen Agricultural University, Aquatic Ecology and Water Quality
Management Group, PO Box 47, 6700 AA Wageningen, The Netherlands}


\begin{abstract}
We show analytically and numerically that the appearance of lumps and gaps in the 
distribution of $n$ competing species along a niche axis is a robust phenomenon 
whenever the finiteness of the niche space is taken into account. 
In this case depending if the niche width of the species $\sigma$ is above or below 
a threshold $\sigma_c$, which for large $n$ coincides with $\frac{2}{n}$, there are 
two different regimes. For $\sigma > \sigma_c$ the lumpy pattern emerges directly from 
the dominant eigenvector of the competition matrix because its corresponding eigenvalue 
becomes negative. For $\sigma \leq \sigma_c$ the lumpy pattern disappears. 
Furthermore,  this {\em clumping transition} exhibits critical slowing down
as $\sigma$ is approached from above. 
We also find that the number of lumps of species vs. $\sigma$ displays a stair-step 
structure. The positions of these steps are distributed according to a power-law.
It is thus straightforward to predict the number of groups that can be packed along a 
niche axis and it coincides with field measurements for a wide range of the model 
parameters.
\end{abstract}


\maketitle

\section{Introduction}

An important problem in ecology is how closely can species be packed in a
natural environment \cite{M74}. A usual way to approach this issue is by
considering the species distributed along a hypothetical one-dimensional niche axis
\cite{M74}. To fix ideas one may consider the niche axis as a 
gradient that is related to the size of organisms.
Each species $i$ is represented by a normal distribution
$P_i$($\xi$)= $\exp$ [-($\xi-\mu_i)^2$/2$\sigma^2$ ] centered at
$\mu_i$, corresponding to its average position $\xi$ on this niche axis,
and with a standard deviation $\sigma$, which measures the width of its
niche.
The competition for finite resources among the $n$ species can be described
by a Lotka-Volterra competition model (LVCM):
\begin{equation}
\frac{dN_i}{dt} = r_i \frac{N_i}{K_i}(K_i-\sum_{j=1}^{n}a_{ij}N_j),
\label{eq:LVCM0}
\end{equation}  
where $N_i$ is the density of species $i$, $r_i$ is its maximum per-capita 
growth rate, $K_i$ is the carrying capacity of species $i$
and the coefficients $a_{ij}$ is the competition coefficient of
species $j$ on species $i$. 
It seems natural to assume that the intensity of the interaction between two
species $i$ and $j$ depends on how close they are along the niche axis. A
measure of this is provided by the {\em niche overlap}, {\i.e.} the 
overlapping between $P_i(\xi)$ and
$P_j(\xi)$. The  competition coefficients $a_{ij}$ can 
be computed by the MacArthur and Levins overlap (MLO) formula \cite{M67}: 
\begin{equation}
a_{ij} =   
\frac{\int_{-\infty}^{\infty}P_i(\xi)P_j(\xi)d\xi}{\int_{-\infty}^{\infty}P_i^2(\xi)}=
\mbox{e}^{-(\frac{\mu_i-\mu_j}{2\sigma})^2}.
\label{eq:ML}
\end{equation}

Recently Scheffer and van Nes \cite{S06} found by simulations that the
combination of LVCM (\ref{eq:LVCM0}) plus MLO (\ref{eq:ML}) yields long transients of lumpy
distributions of species along the niche axis
[For asymptotic times, the lumps are thinned out to single species unless 
a stabilizing mechanism/term is included, as it was shown in \cite{S06}.]

This phenomenon of spontaneous emergence of 
self-organized clusters of look-a-likes separated by gaps with no survivors
was dubbed by the authors as {\em self-organized similarity} (SOS). It was
recognized as an important new finding in an established model in ecology \cite{N06,M07}
In addition, there is empirical evidence for self-organized
coexistence of similar species in communities ranging from mammal
\cite{S99} and bird communities \cite{H92} to lake plankton \cite{H01}.

However, there has been some controversy on whether this lumpy distribution
of species is indeed a robust result or rather depends strongly on details of the
model, like the competition kernel \cite{A08,P07}.

Here we show that the lumpy pattern is a robust phenomenon
provided one takes into account the {\em finiteness} of
the niche axis. Thus, truncation besides being a crucial assumption
which guarantees clustering, allows the analytical computation 
of the eigenvalues and eigenvectors of the competition matrix
{\bf A} with elements $a_{ij}$ given by (\ref{eq:ML}). Furthermore, we show
that ultimately solving the linear problem is enough to get both the
transient pattern
-lumps and gaps between them-   as well as the asymptotic equilibrium.
The plan of this work is as follows:

Since an analytic solution for realistic conditions - species randomly
distributed along a finite and non periodic niche axis, each with a
different
per capita growth rate $r_i$ and carrying capacity $K_i$ - is not possible, 
we will consider in section 2 a series of simplifications. 
We get an analytic expression for the
state of this simpler system, in terms of the dominant eigenvector of {\bf
A}. 
It provides a qualitatively good description of the system for not too
short times and becomes very good for asymptotic times.
Part of the material of this section was presented in a previous
short paper \cite{F09}, but there are some important differences like
considering a
less rough approximation together with some steps better explained. 

In section 3  we show, using simulations, that all these simplifications do
not destroy SOS:  lumps and gaps remain in
the case of a finite linear niche axis no matter if the niche is non
periodic (\ie it has borders), or the species are randomly distributed, or $r$ and $K$
changes from species to species.
Indeed we go further and show that SOS occurs in niches of more than one
dimensions or when interaction kernels different from the Gaussian kernel
are considered.

In section 4 we show that the prediction of the number of lumps as a
function
of $\sigma$ is in good agreement with measures 
in several ecosystems \cite{M74}, provided $\sigma$ is greater than a 
threshold value $\sigma_c$. For this critical value it occurs a bifurcation 
which is responsible for the {\em clumping transition}.

Section 5 is devoted to conclusions and to put our results in its proper 
perspective, addressing some general concerns about SOS and comparing with
other different approaches.

\section{AN ANALYTICAL PROOF OF SELF-ORGANIZED SIMILARITY IN A SIMPLIFIED
CASE}

We start by considering the following simplifications:

\noindent {\it S1 -} The $n$ species are evenly distributed along
a finite niche axis of length $L$ = 1,
{\it i.e.} $\mu_i$ = $(i\!-\!1)/n$ ($i$=1,...,$n$).

\smallskip

\noindent {\it S2 -} To avoid border effects, the niche is defined circular,
i.e.
periodic boundary conditions (PBC) are imposed. 
This is done by just taking the smallest of
$\vert$$\mu_i$-$\mu_j$$\vert$ and 1-$\vert$$\mu_i$-$\mu_j$$\vert$
as the distance between the niche centers. 

\smallskip
  
\noindent {\it S3 -} All species have the same per capita growth rate 
which we take equal to 1, $r_i$ = 1  for all $i$. 

\smallskip

\noindent {\it S4 -} The carrying capacity  $K$ is also homogeneous, 
$K_i$ = $K$  for all $i$. 

\smallskip

Under the simplifying conditions {\it S3} and {\it S4} 
the system of equations (\ref{eq:LVCM0}) reduces to 
\begin{equation}
\frac{dx_i}{dt} = x_i(1-\sum_{j=1}^{n}a_{ij}x_j),
\label{eq:LVCM}
\end{equation}  
where $x_i$ is the density of species $i$, normalized by its carrying
capacity $K_i$ ($x_i$ = $N_i/K_i$).

\begin{figure}[htp]
\begin{center}
  \includegraphics[width=1\textwidth]{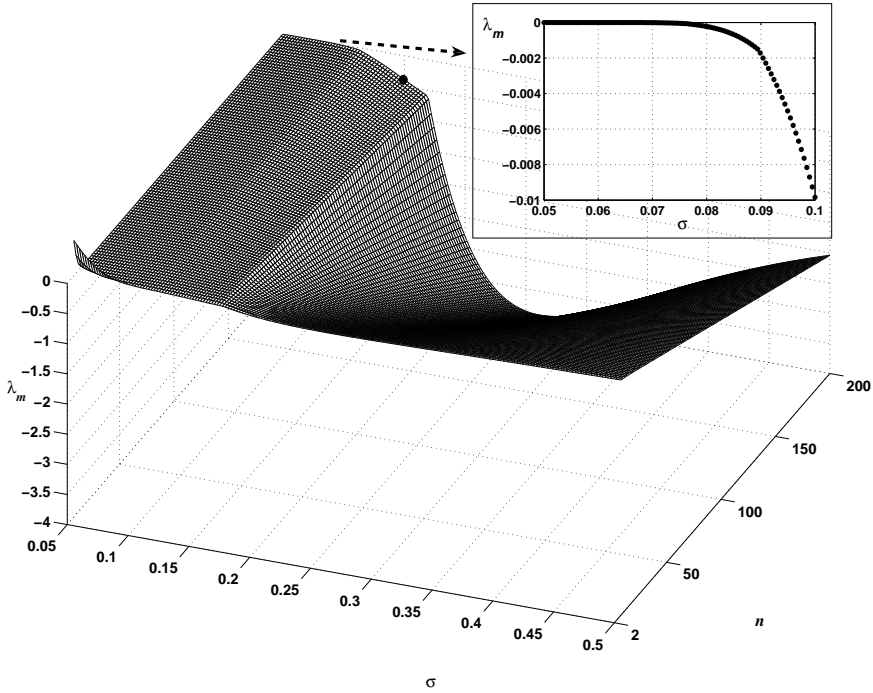}
  \caption{The minimal eigenvalue of {\bf A}, $\lambda_m$ , determined from
equation \ref{eq:eigenvals}, as a function of $n$ and $\sigma$.   The black spot  
denotes the point $n$ = 200 and $\sigma$  = 0.15. Inset: a zoom of
$\lambda_m$ vs. $\sigma$ for $n$ = 200 in the interval 0.05 $\le \sigma \le$
0.1.}     
\label{fig:surf}
\end{center}
\end{figure}

An equilibrium of the system (\ref{eq:LVCM})is specified by a set of
densities $x_i^*$, one for each species $i$, verifying:  
\begin{equation}
x_i^*(1-\sum_{j=1}^{n}a_{ij}x_j^*)=0.
\label{eq:EQ1}
\end{equation}  
A standard procedure to check the stability of this
equilibrium is linear stability analysis. That is, to consider, initially small, 
disturbances $y_i(0)$ from the equilibrium values $x_i^*$ and study their fate $y_i(t)$ as the
time grows. 
Let's take $x_i^* = x^* \forall i$ which, by virtue of conditions {\it S1} and {\it
S2}, is an exact equilibrium \footnote{We later 
checked by simulations that all the derivations below are independent from
the initial condition: the same results are obtained
when starting from a completely random assignation of densities.}.
The evolution equation for $y_i(t)$ can be written as
\begin{equation}
\frac{dy_i}{dt} = -(x^*+y_i(t))\sum_{j=1}^{n}a_{ij}y_j(t).
\label{eq:dy1}
\end{equation}  
Since the coefficients of the matrix {\bf A} given by (\ref{eq:ML}) are
symmetric, in the eigenvector basis $\{ \mbox{{\bf v}}_i\}$, it
becomes diagonal with all its eigenvalues $\lambda_i$ real. Hence 
integrating equation (\ref{eq:dy1}) 
, and using that $y_i(0)$ is small, $y_i(t)$ can be approximated by 
\begin{equation}
y_i(t)\simeq y_i(0)e^{-x^*\lambda_i t}.
\label{eq:y_i} 
\end{equation} 
Thus, for asymptotic times, {\bf y} becomes proportional to the dominant
eigenvector
{\bf v}$^m$, the one associated with the minimum eigenvalue of {\bf A},
$\lambda_m$, {\it i.e.} 
\begin{equation}
\mbox{\bf y}(t) \propto
e^{-x^*\lambda_m t}
\mbox{\bf v}^m \;\;\; \mbox{(for large times)}.
\label{eq:yasympt}
\end{equation}
We will show that, for a wide range of the parameters $n$ and $\sigma$,
$\lambda_m$($n$,$\sigma$) is in general negative (see below). Hence, from
(\ref{eq:yasympt}), {\bf y} is amplified over time
instead of decaying to zero (as it would happen in the case 
of a positive $\lambda_ m$). 
Therefore, for large times, from (\ref{eq:dy1}) 
we can express the time derivative of {\bf x} as 
\begin{equation}
\frac{d\mbox{\bf x}}{dt} = -\mbox{\bf x}(t) \lambda_m \mbox{\bf v}^m,
\end{equation}
and by integration we get the approximated solution given by
\begin{equation}
\mbox{\bf x} \approx
e^{-\lambda_m \mbox{\bf v}^m t} \;\;\; \mbox{(for large
times)}.
\label{eq:xfin}
\end{equation}
Analytic expressions for the eigenvalues and eigenvectors of {\bf A}  are
not
known for the general case of random distributions of species on a niche
axis with arbitrary boundary conditions. However, for the simpler case when 
the $n$ species are evenly spaced along the niche axis, $\mu_j$ =
$(j\!-\!1)/n$ 
(with the index $j$=1,...,$n$), and PBC (the simplifying conditions {\it S1}
and {\it S2}) 
{\bf A} becomes a matrix whose rows are cyclic permutations of the first
one:
\[
\begin{bmatrix}
c_1 & c_2 & \dots & c_{n-1} & c_n \\ 
c_n & c_1 & \dots & c_{n-2}  & c_{n-1} \\ 
\dots & \dots & \dots & \dots & \dots \\ 
\dots & \dots & \dots & \dots & \dots \\
c_2 & c_3 & \dots & c_n  & c_1 
\end{bmatrix}
\]
with $c_j(n,\sigma)= e^{-(\frac{\widetilde{j-1} }{2 \sigma n})^2}$, where 
the tilde stands for $\pmod{\frac{n+2}{2}}$ implementing then PBC.  
For this case, the eigenvalues $\lambda_k$ and the components 
of the eigenvectors $v^k$ ($k$ = 1,...,$n$) are given by \cite{B52}:
$$\lambda_k  = \sum_{j=1}^{n}c_j(n,\sigma)e^{i 2 \pi (k-1)\mu_j}$$
\begin{equation}
=  \sum_{j=1}^{n}c_j(n,\sigma)e^{i 2 \pi (k-1)(j-1)/n}, 
\label{eq:eigenvals}
\end{equation}
and
$$v_j^k  = n^{-\frac{1}{2}} [ \cos (2\pi(k\!-\!1)\mu_j) + \sin
(2\pi(k\!-\!1)\mu_j) ]  $$
\begin{equation}
= n^{-\frac{1}{2}} \left[ \cos \left( \frac{2\pi(k\!-\!1)(j\!-\!1)}{n}
\right) + 
\sin \left( \frac{2\pi(k\!-\!1)(j\!-\!1)}{n} \right) \right].  
\label{eq:eigenvects}
\end{equation}
Since the matrix {\bf A} is symmetric $c_j=c_{n+2-j}$. Therefore, 
from (\ref{eq:eigenvals}) one can see that the eigenvalues occur in pairs: 
$\lambda_k = \lambda_{n-k}$, with the exception of $\lambda_1$  (and of
$\lambda_{n/2+1}$ if $n$ is even). Furthermore, these paired eigenvalues can
be expressed as
\begin{equation}
\lambda_k = 2\sum_{j=2}^{n/2}c_j(n,\sigma)\cos[2\pi(k-1)(j-1)/n]. \\
\label{eq:eigenvals2}  
\end{equation}
Equation (\ref{eq:eigenvals}) can be used to determine
the index $k$ = $m$ that gives the minimal eigenvalue, for $n$ and
$\sigma$ given, $\lambda_m(n,\sigma)$ (as we have just
seen, the index $k$ = $n$-$m$+2 produces the same value). 
The surface depicted in Fig. \ref{fig:surf} corresponds to $\lambda_
m$($n$,$\sigma$) computed for a grid 2$\leq$ $n$ $\leq$ 200, and 0.05
$\leq$ $\sigma$ $\leq$ 0.5. Notice that $\lambda_ m$ is negative except for
small values of $\sigma$ and becomes positive when $n$ $<$ 8. 
The substitution of the dominant eigenvector {\bf v}$^m$, which from
(\ref{eq:eigenvects})
has $m$-1 peaks and $m$-1 valleys, into (\ref{eq:xfin}) allows to predict
the distribution of species for long enough times. 

\begin{figure}[htp]
\begin{center}
  \includegraphics[height=11cm]{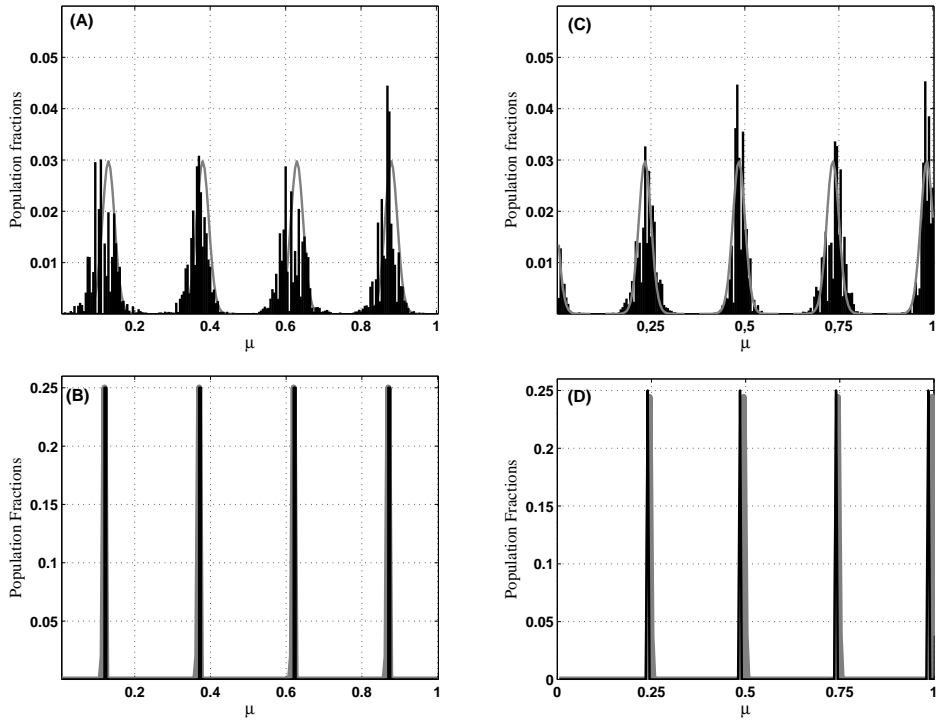}
   \caption{Population fractions $\hat{x}_i$ for $n$ = 200 and $\sigma$  = 0.15. In black results
from a simulation after $t$ generations and in gray $\exp$ [
-$\lambda_m${\bf v}$^m$ $t$]. (A) and (B): Species
 evenly spaced along the niche axis for $t$ = 1000 and $t$ = 10,000
 generations, respectively. (C) and (D): Species randomly distributed along
 the niche axis
for $t$ = 1000 and $t$ = 10,000. 
}     
\label{fig:Even_and_Rand}
\end{center}
\end{figure}

The results we got were checked by numerical simulations. 
In these simulations the initial values for the $x_i$ are random numbers
between 
0 and 1. Then the system of differential equations (ODE) is integrated for a 
given final time.
In Fig. \ref{fig:Even_and_Rand} we compare this analytical approximation
with simulations. 
For instance, if $n$ = 200 and $\sigma$ = 0.15 we
get  $m$ = 5  ( and $m$ = 200-5+2 = 197), $\lambda_ m$ = 0.3938 and the
components of {\bf v}$^m$ are given by  
$\sqrt{\frac{1}{n}}\sin [8\pi\mu_j]$ + $\sqrt{\frac{1}{n}}\cos [8\pi\mu_j]$.
Panel (A) of Fig.\ref{fig:Even_and_Rand} if for $t$ =1000
generations. The agreement is quite good and the quality of the
agreement improves with time, until it becomes very good when the
lumps are thinned to single lines as it is shown in panel (B)\footnote{The
gray lines, generated from {\bf v}$^m$, are actually lines. They were drawn 
tick just to show their coincidence with the black thin lines produced by
simulations.}. This happens because we are not considering any lump
stabilizing term like the one considered in \cite{S06}. 
Notice that ultimately the lumps and gaps coincide,
respectively, with the $m$ -1 maximums and minimums of {\bf v}$^m$. 
The integer $m$, which gives the minimal eigenvalue, is a function of the
width $\sigma$  of the niche, $m$ = $m$($\sigma$). It does not depend from
$n$ provided $n$ is large enough. Nevertheless, as we will show in section
IV, $m$ becomes a function of $n$ and $\sigma$ for small values of both
these parameters.    
For example, for $\sigma$=0.15, $m$-1 = 4 for all even $n$ greater or equal
than 8. This lower $n$ limit arises because the maximum possible number of
peaks 
that can be accommodated with $n$ vector components is $n$/2 
(one half of the components of {\bf v}$^m$ pointing up and the other half
down). So in this particular case $n$/2 must be greater or equal than 4,
and, in
general,  $n$/2 must be greater or equal than $m$-1.

Another remarkable result about $m$ is that it is always an odd number 
(and then the number of clumps is even). The reason for this
can be traced from the cosines appearing in (\ref{eq:eigenvals2}) making
contributions to the eigenvalues of opposite signs: positive for odd $k$ and
negative for even $k$. As a consequence the number of peaks, equal to $m$-1,
is always even.

\section{SELF-ORGANIZED SIMILARITY PERSISTS UNDER MORE REALISTIC
ASSUMPTIONS}

In order to consider more realistic assumptions, abandoning the simplifying 
conditions {\it S1-S4}, we rely in the following exclusively on simulations. 
Since the emphasis in SOS is on transient maintenance of clumps of similar
species, one  might wonder about how initial conditions determine the
results,
and how species that are being driven extinct ever managed to get up to high
density in the first place. 
So, as before, the ODE system is integrated starting from initial $x_i$
which 
are random numbers between 0 and 1. We checked in all the cases that changes
in the initial populations don't introduce qualitative changes.

{\it 1. From evenly to randomly distributed species.}

\vspace{2mm}

What happens in the general case of randomly distributed species over the
niche axis? 
In this case the spectrum and {\bf v}$^m$ are obtained numerically from 
{\bf A}. 
It turns out that simulations produce quite the same results. 
We illustrate this in Fig.\ref{fig:Even_and_Rand}
where we plot the population fractions normalized to one,
$\hat{x}_i=\frac{x_i}{\sum_{i=1}^{n} x_i}$, 
for the particular parameter values $n$=200 and $\sigma$=0.15. 
The resemblance is clear when 
comparing  panels (C) and (D) with, respectively, (A) and (B).
In fact, the spectrum of eigenvalues
in both cases is very similar as it is shown in Table 1 
(the values on the right correspond to averages among simulations).

\begin{center}
\begin{tabular}{|c|c|c|}
\hline
 & Evenly spaced & Randomly distributed          \\
\hline
$\lambda_1$   & -0.3938  & -0.4 $\pm$0.01        \\
$\lambda_2$   & -0.3938  & -0.4 $\pm$0.01        \\
 \vdots  &  \vdots       &  \vdots               \\
$\lambda_{198th}$ & 45.391 & 46$\pm$0.96         \\
$\lambda_{199th}$ & 45.391 & 46$\pm$0.96         \\
$\lambda_{200th}$ & 104.387 & 105 $\pm$1.93      \\
\hline
\end{tabular}
\end{center} 
\begin{center}
Table 1: eigenvalues of {\bf A} for $n$ = 200 \& $\sigma$ = 0.15 ordered
from small to large \ie $\lambda_m=\lambda_1=\lambda_2$.
\end{center}

{\it 2. Taking into account border effects in a linear niche axis.}

\vspace{2mm}

\begin{figure}[htp]
\begin{center}
     \includegraphics[height=11cm]{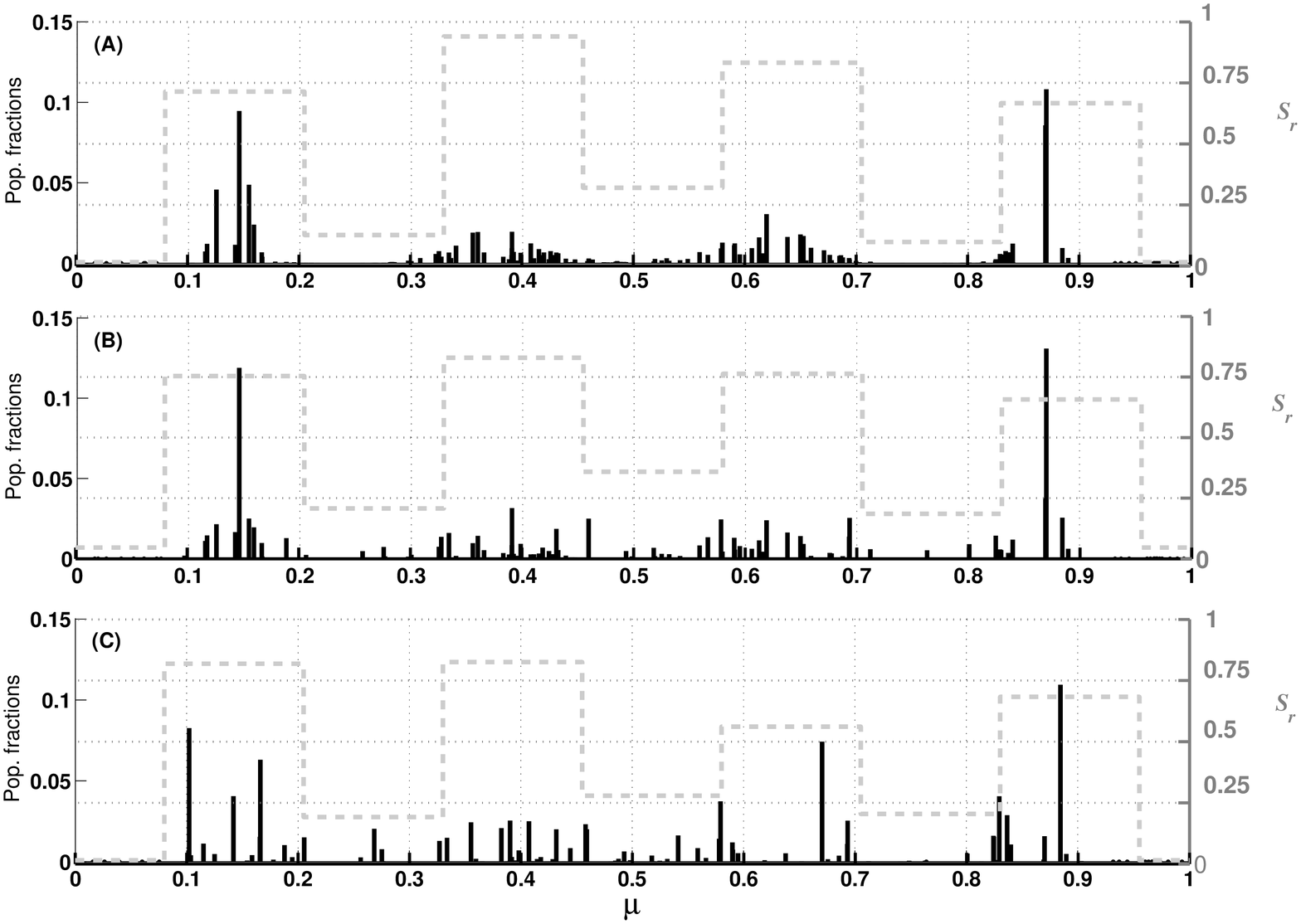}
    \caption{      
Fraction of species $\hat{x}_i$ (black bars, left vertical axis) for $n$ = 200, $\sigma$  = 0.1 and 
open boundary conditions (coefficients given by (\ref{eq:OBC}) with $L$=1) 
after 500 generations and the corresponding entropy for each lump and gap region $S_r$
(gray dashed lines, right vertical axis). (A): Uniform maximum growth rate $r$ and carrying capacity $K$. 
(B): Varying $r$ (the $r_i$ are random numbers 
with average value equal to 1) and uniform $K$.
(C): Varying $r$ (the $r_i$ are random numbers 
with average value equal to 1)  
and $K$ ( $\delta K_{max}/\bar{K}$ = 0.2, see text)
from species to species.}     
\label{fig:x3}
\end{center}
\end{figure}

We also analyzed what happens when a {\em linear}, instead of a circular
niche 
axis (PBC), of length $L$ is considered.
The competition coefficients for these open boundary conditions (OBC)
are now given by
\begin{equation}
a_{ij} =   
\mbox{e}^{-(\frac{\mu_i-\mu_j}{2\sigma})^2}
\frac{\mbox{erf} ( \frac{2L\!-\!\mu_i\!-\!\mu_j}{2\sigma} )+
\mbox{erf} ( \frac{\mu_i\!+\!\mu_j}{2\sigma} )}
{\mbox{erf} (\frac{L\!-\!\mu_i}{\sigma} )+\mbox{erf}(\frac{\mu_i}{\sigma})}.
\label{eq:OBC}
\end{equation}
When using competition coefficients given by (\ref{eq:OBC}) with $L$=1, 
again, a lumpy pattern emerges although it shows some quantitative
differences. For example, a four lump pattern occurs for smaller values of 
$\sigma$, e.g. $\sigma$= 0.12 instead of $\sigma$= 0.15 
(panel (A) in Fig.\ref{fig:x3}). 
Additionally, although $\lambda_m$ is still negative,
due to the factor multiplying the Gaussian in (\ref{eq:OBC}), 
the matrix {\bf A} is no longer symmetric and so there appear complex
eigenvalues. 

\vspace{2mm}
{\it 3. The effect of a non uniform growth rate.}
\vspace{2mm}

Simplification {\it S3} was to consider a uniform $r$.
Indeed it is simple to realize that an $r$ varying from species to species 
does not introduce major changes. This is because what is relevant for the 
equilibrium values $x_i^*$ are the terms between brackets in the LVCM
equations (see panel (B) in Fig.\ref{fig:x3}).

\vspace{2mm}
{\it 4. The effect of the heterogeneity in the carrying capacity.}
\vspace{2mm}

We find that when variations $\pm$$\delta$$K_i$ of the carrying capacity
around
an average value $\bar{K}$ occur in such a way that the amplitude of these
fluctuations, $\delta$$K_{max}$, is no greater than 10 \% of $\bar{K}$ the
lumpy 
pattern changes but is similar to the one corresponding to the homogeneous
case. If, in addition, one assumes that the carrying capacity of neighbor
species along the niche axis have similar carrying capacities and
larger variations are only possible for species which are far away on the
niche axis, then larger values of $\delta$$K_{max}$/$\bar{K}$ still preserve
SOS (panel (C) in Fig.\ref{fig:x3}).  
On the other hand, strong random variations of the carrying
capacity along the niche axis in general destroy the SOS pattern.

In Fig.\ref{fig:x3} we show the population fractions 
obtained when the more realistic conditions 2 to 4 
are gradually taken into account. 
In the three panels we plot the results produced by simulations starting
from the same initial distribution of populations.  
Panel (A) corresponds
to OBC and homogeneous $r$ and $K$, panel (B)
to OBC, heterogeneous $r$ and homogeneous $K$
and panel (C) to OBC and heterogeneous $r$ as well as $K$.
Notice that although the lumpy structure becomes less 
clear as the original restrictions are lifted, it is still recognizable 
in panel (C).
In order to provide a more quantitative test for the clumping to the 
favorable niches, it is necessary to introduce an observable which measures
species coexistence or diversity. 
Among the different indices proposed to measure species diversity
perhaps the most common is the Shannon-Wiener index \cite{P69},\cite{H73}, 
or in the physics language the well known entropy $S$, defined by

$$S = -\sum_{i=1}^{n} \hat{x}_i \ln \hat{x}_i.$$

Moreover, entropy analysis has been used 
to quantify species diversity and niche breadth \cite{E77} and to recognize
ecological structures (see \cite{P88} and references therein).
Therefore we proceed as follows. From the homogeneous $r$ and $K$ situation
we obtain the modulation along the niche axis determining the number and
positions of lumps and gaps. In this specific case there are four lumps 
separated by gaps all of the same length. 
Thus we divide the niche axis into 9 regions: 4 lumps and 3 gaps between
them, all the 7 of length 0.125, plus the two smaller gaps at the niche borders
completing the remaining length of 0.125. 
The amount of entropy $S_r$ calculated for each region 
($r$=1,2,...), measures the species diversity (represented by gray
dashed lines in panels (A)-(C)). Notice that the profile of
$S_r$ for the three situations is similar although, as expected, 
it offers more clear cut evidence of lumps and gaps for 
the homogeneous situation of panel (A): the entropy is in general lower
(higher) at the gaps (lumps of coexistence) than in panels (B) or (C). 
Therefore, we conclude that the considered simplifications don't introduce
substantial changes and that SOS survives in more realistic conditions.

\vspace{2mm}

{\it Other competition kernels and multidimensional niches}
\vspace{2mm}

\begin{figure}[htp]
\begin{center}
     \includegraphics[height=9cm]{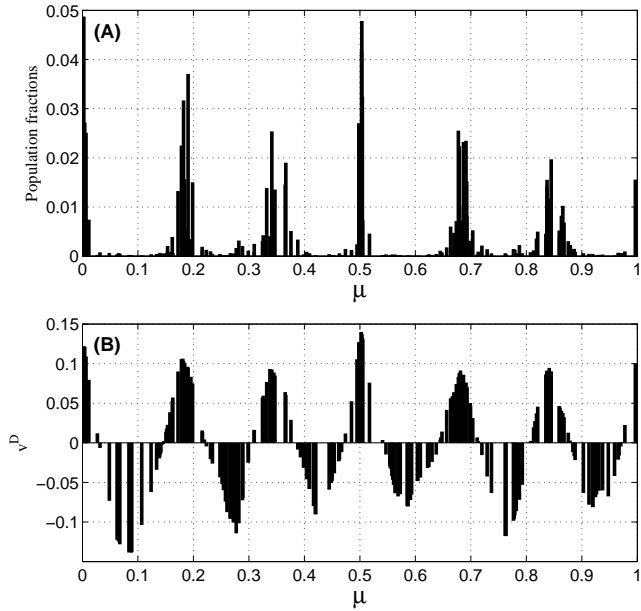}
    \caption{Results for a non-Gaussian kernel 
with $p$=1.5, $n$ = 200, $\sigma$  = 0.19 and PBC.       
(A): Distribution of species for after 2500 generations.
(B): Components of the dominant eigenvector {\bf v}$^m$.
}     
\label{fig:p15}
\end{center}
\end{figure}

\begin{figure}[htp]
\begin{center}
     \includegraphics[height=11cm]{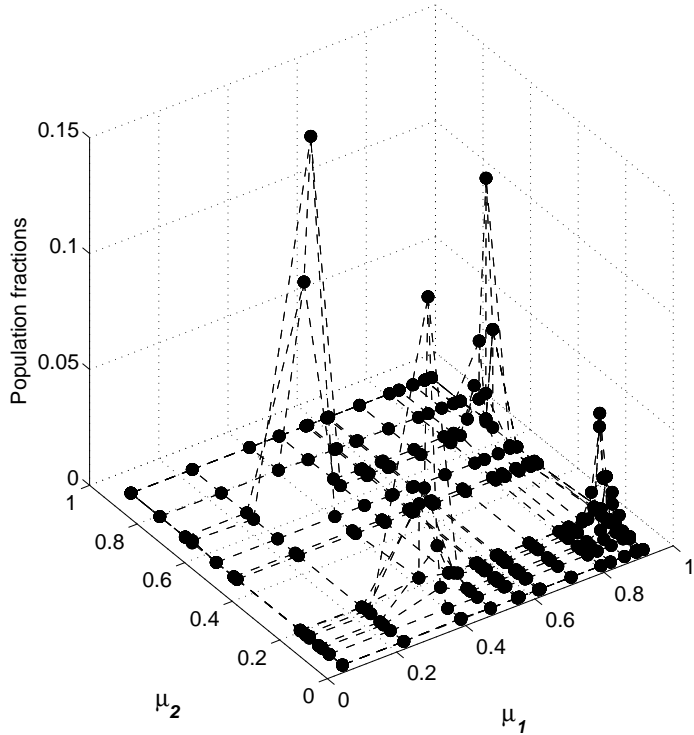}
    \caption{ Distribution of species in a two-dimensional niche of coordinate
 axis $\mu_1$ and $\mu_2$ for $n$ = 15$\times$15= 225 species, 
 and $\sigma$  = 0.2, PBC after 100 generations. 2$\times$2=4 clumps
are observable.}     
\label{fig:multi}
\end{center}   
\end{figure}

It was argued that the formula (\ref{eq:ML}) is a special case and 
that competition coefficients are typically non-Gaussian \cite{A08,W75}. 
Some recent analyses 
explore more general non-Gaussian competition kernels of the form
\cite{D07}:
\begin{equation}
a_{ij} =   
\mbox{e}^{-(\frac{\mu_i-\mu_j}{2\sigma})^p},
\label{eq:NonGaussian}
\end{equation}  
which reduces to the Gaussian one for $p$=2. 
Moreover, it was claimed that Gaussian competition does not lead to patterns
but 
is a borderline case between patterns and non-patterns regimes \cite{P07}.
However, this depends on whether
or not one takes into account the finiteness of the niche axis. 
When it is taken into account, as we do by using a  
truncated kernel, $p$=2 is no longer a border case. 
Rather the lumpy pattern occurs
for any real kernel exponent $p$ above 1, for example, $p$ = 1.5 as is it is
shown in panel (A) of Fig. \ref{fig:p15} for $\sigma$=0.19.
This is because the only change in the formula for the eigenvalues
(\ref{eq:eigenvals}) 
is in the coefficients $c_j(n,\sigma)$ which, for a general value of the
exponent
$p$,  are given by
$c_j(n,\sigma)= e^{-(\frac{j-1}{2 \sigma n})^p}$ while the expression 
(\ref{eq:eigenvects}) for the eigenvectors remains unchanged.
Panel (B) of this figure is a plot of the components of the {\bf v}$^m$
showing that its peaks (valleys) coincide with the lumps (gaps).
Another common criticism is that it is not very realistic to consider 
a one-dimensional niche, rather  
niches (utilizations) in general are 
multi-dimensional \cite{S74,P82}
It turns out that a multi-dimensional niche only 
makes the math a little bit less straightforward.
Suppose that the $n$ species are distributed at random in a 2-dimensional
niche with axes $\mu_1$ and $\mu_2$. Then one can assign 
an index $i$ to each population, located at the point in niche space given
by 
a couple ($\mu_{1i}$,$\mu_{2i}$), and group them into a vector of $n$
components. Therefore, the expression for the competition 
coefficient between species $i$, located in this niche space at a point
of coordinates ($\mu_{1i}$,$\mu_{2i}$), and 
species $j$, at ($\mu_{1j}$,$\mu_{2j}$), can be written as
\begin{equation}
a_{ij} =   
\mbox{e}^{-\frac{(\mu_{1i}-\mu_{1j})^2+(\mu_{2i}-\mu_{2j})^2}{(2\sigma)^2}}.
\label{eq:2dN}
\end{equation}
It turns out that this preserves the {\it cyclic} property of the {\bf A} 
matrix - its rows are cyclic permutations of the first one -, 
a property required to
get the expressions for the  
eigenvalues (\ref{eq:eigenvals}) and the eigenvectors
(\ref{eq:eigenvects}) \cite{B52}.
Fig. \ref{fig:multi} shows the results for $\sigma$=0.2. 
It shows a general result we found: if in the case of a one-dimensional 
niche ($d$=1), for a given value of $\sigma$, there are $m$-1 lumps,
for a two-dimensional niche ($d$=2) there occur ($m$-1)$\times$($m$-1) lumps
(for $\sigma$ = 0.2 there are 2 lumps for $d$=1 while for $d$=2 there are 
4 lumps). 

\section{THE DEPENDENCE OF CLUMPING ON THE NICHE WIDTH AND THE CLUMPING
TRANSITION}

\begin{figure}[htp]
\begin{center}
     \includegraphics[height=10cm]{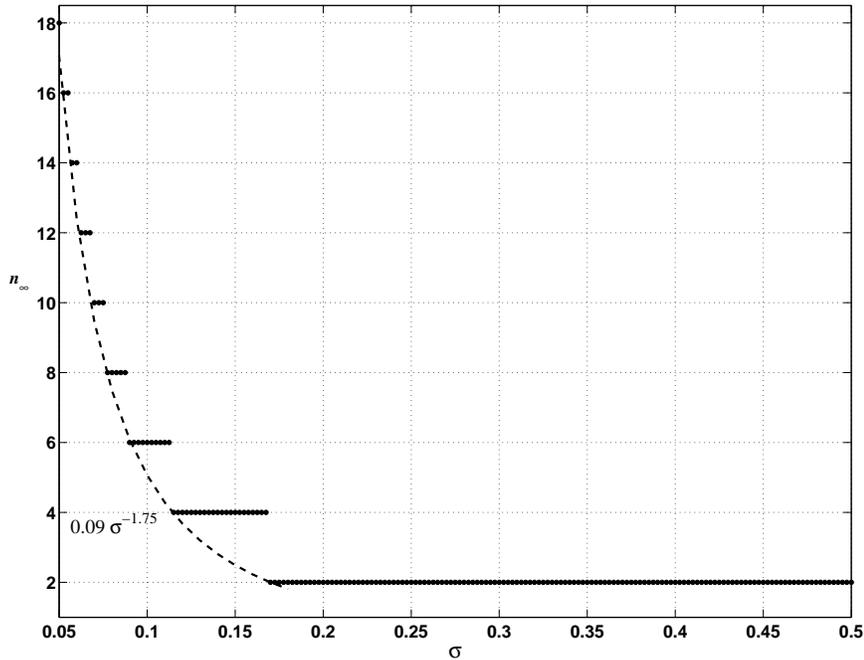}
    \caption{Number peaks $n_\infty$ of the dominant eigenvector {\bf v}$^m$ 
as a function of $\sigma$ ($n$ = 200). 
The jumps follow  a power-law distribution 
indicated by a dashed line.}     
\label{fig:Steps}
\end{center}   
\end{figure}

\begin{figure}[htp]
\begin{center}
     \includegraphics[height=9cm]{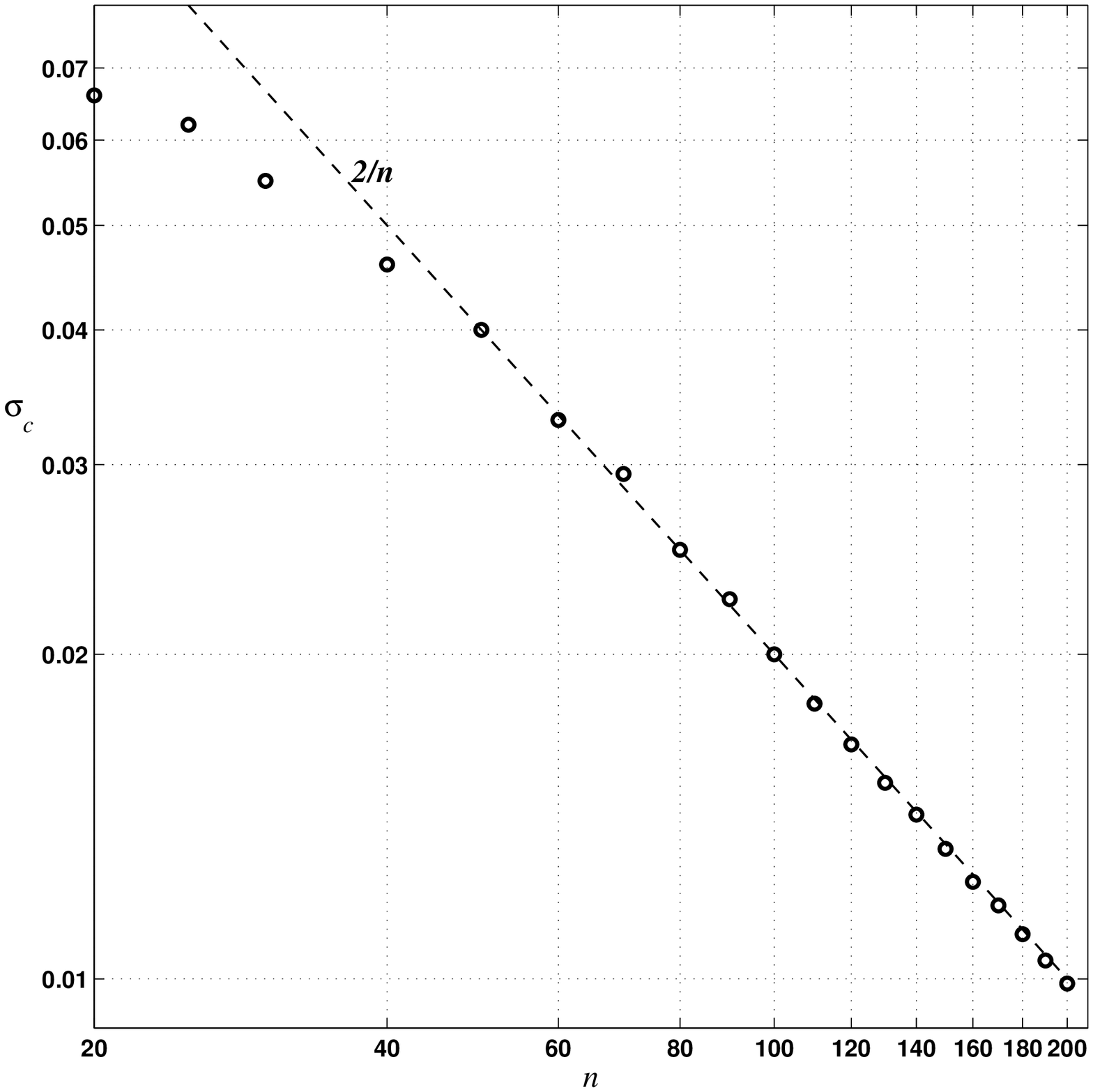}
   \end{center}
    \caption{Log-log plot of $\sigma_c$ vs. $n$ for $n$ between 20 and 200. 
The dashed line corresponds to 2/$n$.}     
\label{fig:sigma_c}
\end{figure}

How close species can be packed along the niche axis
is commonly measured by the parameter $d/\sigma$, where $d$ is the
separation 
between species \cite{M74}. 
In the case of the model under consideration,  
$d$ can either measure the separation between a) lumped groups of species,
persisting during long transients or b) surviving species (one per lump),
for asymptotic times.
So we get either an estimate for the species packing or for the group packing. 
In any event, this distance coincides with the inverse of the number of
peaks of {\bf v}$^m$, which depends on $\sigma$,
and is given by $n_\infty$ ($\sigma$) = $m$($\sigma$)-1.
Fig.\ref{fig:Steps} shows $n_\infty$ for $\sigma$ ranging from 0.05 to 0.5 
and the number of species fixed to $n$ = 200.
There is a series of steps, located at values $\sigma_s$, 
that become wider as $\sigma_s$ increases. The height of these steps is
always 2.
This is because, as we have seen in section 2, the number of peaks of {\bf
v}$^m$ 
is always an even number. 
That is, if $\sigma_s^-$ ($\sigma_s^+$) corresponds to $\sigma$ tending to 
$\sigma_s$ by the left (right), then 
$n_\infty$($\sigma_s^-$) = $n_\infty$($\sigma_s^+$)+2. For example,  
if $\sigma$ $>$ $\sigma_s$ $\simeq$ 0.169 then {\bf v}$^m$ has  
always 2 peaks, below this value the 
number of its peaks jumps to 4, and so on. We find that
when $\sigma$ tends to $\sigma_s^+$ , $n_\infty$ can be fitted with the
power-law 
0.09$\sigma_s^{-1.75}$ (dashed line in Fig.\ref{fig:Steps}). 
The packing parameter in the different step regions can be approximated, by
taking the number of peaks at each $\sigma_s$ as the semi-sum of 
the numbers of peaks at each side of the step, as: 
\begin{equation}
\frac{d}{\sigma_s}\simeq \frac{1}{\sigma_s 1/2 (n_\infty(\sigma_s^+) +
n_\infty(\sigma_s^-))} 
= \frac{1}{\sigma_s+0.09\sigma_s^{-0.75}}.
\label{eq:d-s}
\end{equation} 
This quotient varies from approximately 1.96 for the first step, at
$\sigma_s$
$\simeq$0.169, to 1.1 for the last step, at  $\sigma_s$ $\simeq$  0.05.
This is in good agreement with many field measurements that found
a species packing ratio always lying between 1 and 2 \cite{M74}. 
It is worth remarking that when $\sigma$ decreases, 
$\lambda_m$ -which is in general negative- increases, 
until at some critical value, $\sigma_c$, it becomes 0.
That is, in Thom's catastrophe theory language \cite{T75}, 
a {\em degenerate critical point} or a {\em Non-Morse critical point}.   
This $\sigma_c$ depends on the number 
of species: it decreases with $n$.
We computed $\sigma_c(n)$ as the values such that $\lambda_m(n,\sigma_c)$ 
becomes 0.
In Fig.\ref{fig:sigma_c} we show this. Notice that for $n \geq 40$ 
$\sigma_c$ scales as 2$n^{-1}$, {\it i.e.} the double of the initial average
separation between species.
As $\sigma$ is decreased in simulations -for a fixed value of $n$- so that
it becomes closer and closer
to $\sigma_c$ and $\lambda_m$ moves towards 0, we observed that 
the time to reach the lumpy pattern grows unbounded.
This is the well known phenomenon of {\it critical slowing down}\cite{G81}
: the characteristic relaxation time of the 
dominant eigenmode is proportional to 1/$\lambda_m$.
In fact, taking $n$=200, for $\sigma$ = 0.15 the 4 lumps are noticeable
after 
typically 500 generations while the 6 lumps
for $\sigma$  = 0.1 require around 20,000 generations 
and for $\sigma$ = 0.075 it takes a huge number of generations (more than
500,000) to produce the 10 lumps pattern. 
The inset of Fig.\ref{fig:surf} is a zoom of $\lambda_m$ vs.
$\sigma$. It shows that, at least for all practical purposes, 
the clumping becomes noticeable at $\sigma \simeq$ 0.075.

\section{CONCLUSIONS AND FINAL COMMENTS}

A realistic mathematical description of the dynamics 
of a large number of species placed along a resource spectrum 
is a complicated issue for which an exact solution is not available.
In fact, analytical work looks at the long-term equilibria of models.
The alternative to deal with the transients are simulations.
    
However a simulation approach, like the one used by Scheffer and van Nes
\cite{S06} 
may leave room for doubts on whether things 
might be artifacts.
We made a series of simplifications which allow an
analytic proof, by working directly with the community
matrix {\bf A}, of the emergence of SOS. Roughly,   
the lumpy pattern one is seeing is the exponential of the dominant 
eigenvector {\bf v}$^m$ of {\bf A} (multiplied by the time).

We later have shown that this is indeed a robust result. 
The clumping phenomenon
does not depend on the boundary conditions nor on
the kernel exponent $p$ (provided it is greater than one), it is quite
independent on the heterogeneity of species parameters $r$ and $K$,   
it occurs for a wide range of $\sigma$, and in more than one niche 
dimensions. 
In addition, Roelke and Eldridge \cite{R08} 
found similar patterns in
a different resource competition model. They also suggest that this
mechanism is not very fragile. 
Additional supporting evidences can be found in \cite{H07}.

A crucial element to get clumping is to take into account the
finiteness of the niche axis. This, besides being
realistic, leads to a $\lambda_ m$ with a negative real part (and then to 
lumps and gaps). We want to remark that either 
OBC or the standard implementation of PBC imply a niche which is finite.
This explains remarkable differences with the outcomes reported in
\cite{P07}.
The procedure they use to implement PBC 
consists in taking a periodic array of copies of the same system.
This ``perfectly periodic''  boundary conditions 
mimic an {\em infinite} niche axis.
The first of such differences is that, in our case, the SOS is robust
against 
variations on the kernel: a negative $\lambda_ m$ is obtained 
whenever the exponent $p$ of the kernel interaction is a {\em real} 
number greater than 1. 
The second difference, is that the parameter $\sigma$, controlling the 
width of each species distribution, plays a fundamental role. 
There is a critical value, $\sigma_c$, below which there is no 
clustering.
On the other hand, in a virtually infinite niche axis, 
since it is always possible to set $\sigma$ = 1 by rescaling $\mu$, 
the clustering should not depend on $\sigma$.
However it seems natural that things in ecosystems depend on $\sigma$, 
and usually the interest is precisely in measuring this effect.
This is another powerful reason to prefer an implementation of
boundary conditions like the one we are using.

Similar approaches to determine pattern
formation in phenotype space have also been used by
Levin and Segel \cite{L85}, Sasaki \cite{S97} and more recently by
Mesz\'ena and co-workers \cite{B09},\cite{Sz06}.
The main difference of our approach is that a discrete set of phenotypes is
considered, instead of a continuum.
This is an important feature, since firstly it can affect the
assessment concerning how robust SOS is. That is, in the case of a 
continuous set of phenotypes, it was shown
that an arbitrarily small perturbation can destroy the continuous
coexistence 
making the species distribution discrete \cite{B09},\cite{Sz06}.
And this could be taken as an evidence of the breaking of SOS.
On the other hand, in our model, from the very beginning, the distribution
of species is discrete and SOS it is rather understood as a strong overlap
of the species
distributions. 
Secondly, it allows to go beyond the large 
$n$ limit and applying it to real communities, involving a number of species 
$n$ of intermediate size. 
For example, this was tested for the case of phytoplankton communities in a lake 
ecosystem involving between 50 to 100 species and the
agreement between theory and empirical data is quite good \cite{FS10}. 
Moreover, in this real ecosystem the fact that 
$\sigma$ is not the same for all species, rather varies from species to
species, doesn't spoil the lumpy pattern. 

There are interesting parallels with similar phenomena in physical
systems: 

\begin{itemize}

\item For example, the fact that the emergence of the lumpy pattern is 
related to the eigenvector with the minimum eigenvalue and the number of 
lumps is determined mostly by the model parameter $\sigma$ resembles the spinodal
decomposition \footnote{In fact we are grateful to one of the anonymous referees 
who pointed out this.}
That is, under the spinodal decomposition 
the system develops a spatially modulated order parameter whose amplitude  
grows continuously from zero and extend throughout the entire system. This
results in domains of a characteristic length scale
called the spinodal length $\lambda_{sp}$ which usually depends strongly 
on temperature (because the second derivative of the free energy 
becomes increasingly negative deep inside the region delimited by the
spinodal) \cite{P06}.

\item The emergence of power laws and critical slowing down
are attractive ingredients to physicists since they are signatures of 
self-organized criticality (SOC) \cite{B96}. 
This tendency to spontaneously 
self-organize into a critical state, without any significant ``tuning'' of
some control parameter, usually reflects a share of the same fundamental
dynamics for many different systems referred to as universality.
While the origin of critical slowing down is clear explained by the
existence of a degenerate critical point, the power law distribution 
for the plateaus  of the number of lumps vs. $\sigma$   
is not completely understood and deserves further analysis.

\item The application of techniques and concepts of statistical
mechanics into very different realms like ecosystems might be of interest to ecologists,
statistical physicists and to the growing community on the intersection of both fields.
In that sense, the analytical proof of clumping is 
based on statistical mechanics results from Berlin and Kac \cite{B52} when 
they were analysing the spherical model of a ferromagnet.

\item 
The calculation of the entropy for different regions of a system, 
to get an overview about the level of correlation between elements in each
region has been used in several contexts closer to physics. 
For example: cellular automata \cite{M94}, 
deterministic models of nonlinear dynamics \cite{D00}, glass-forming
materials \cite{Mu89}, astrophysics of galaxies and clusters \cite{Vo03},
image processing \cite{Ba07}, to mention some.
In our case it has shown to be useful to identify the lumpy structure.

\end{itemize}

To conclude, it is remarkable that the predictions on the number of groups of 
species that can be packed along the niche axis
are quantitatively consistent with field data for a wide range of values
of both the width of the niche and the number of species.

\ack{H.F thanks Raúl Donangelo for his thorough criticism and
suggestions and acknowledge financial support from PEDECIBA and ANII Uruguay}

\section*{References}
\bibliographystyle{unsrt}
\bibliography{bibl}

\end{document}